\documentclass[aps,twocolumn,float,prd,psfig]{revtex4}
\input epsf
\usepackage{graphicx}
\newcommand{\beq}{\begin{equation}}
\newcommand{\beqa}{\begin{eqnarray}}
\newcommand{\eeq}{\end{equation}}
\newcommand{\eeqa}{\end{eqnarray}}

\begin{document}
\title{
Second and higher-order quasi-normal modes
in binary black hole mergers
} 
\author{Kunihito Ioka$^{1}$ and Hiroyuki Nakano$^{2}$
}
\affiliation{$^{1}$
Department of Physics, Kyoto University, Kyoto 606-8502, Japan \\
$^{2}$Center for Computational Relativity and Gravitation, 
School of Mathematical Sciences, \\
Rochester Institute of Technology, 
Rochester, New York 14623, USA
}

\begin{abstract}
 Black hole (BH) oscillations known as quasi-normal modes (QNMs) are one
 of the most important gravitational wave (GW) sources. We propose that
 higher perturbative order of QNMs, generated by nonlinear 
 gravitational interaction near the BHs, are detectable and worth
 searching for in observations and simulations of binary BH mergers. 
 We calculate the
 metric perturbations to second-order and explicitly regularize the
 master equation at the horizon and spatial infinity. 
 We find that the
 second-order QNMs have frequencies twice the first-order ones and the
 GW amplitude is up to $\sim 10\%$ that of the first-order one. 
 The QNM frequency would also shift blueward up to $\sim 1\%$.
 This provides a new test of general relativity as well as 
 a possible distance indicator.
\end{abstract}
\maketitle

\section{Introduction}
Direct detections of gravitational waves (GWs)
will become a reality in the near future with current and future
detectors, such as LIGO, LISA and DECIGO/BBO~\cite{Seto:2001qf}. 
GWs can not only provide a test of general relativity but also open 
a new window on the universe.

One of the most important GW sources are the quasi-normal modes (QNMs)
of black holes (BHs)~\cite{Kokkotas:1999bd}. QNMs are oscillations
of the BH metric perturbations, and they are damped by emitting GWs. QNM
frequencies are complex with the real part representing the oscillation
frequency and the imaginary part representing the damping. 
By observing the QNM frequencies, one can determine the mass and angular
momentum of spinning BHs.

The most promising sources that excite QNMs are binary BH mergers. For
these events QNMs will be detected with high signal-to-noise ratio (SNR)
(e.g., SNR $\sim 10^5$ for $\sim 10^8 M_{\odot}$ BH mergers at $\sim
1$Gpc by LISA)~\cite{Flanagan:1997sx} since a large fraction of energy
($E \sim 1\% \times M$ for equal mass mergers with a total mass $M$) 
goes into GWs. The merger rate is also estimated to be
large enough~\cite{Enoki:2004ew,Ioka:2005pm}.
In numerical relativity there has been recently breakthroughs
for calculating the entire phase of binary BH 
mergers~\cite{Pretorius:2005gq}. 
The result is that the $\ell=2$, $m=2$ QNM is dominant, carrying away 
$\sim1\%$ of the initial rest mass of 
the system~\cite{Berti:2007fi}.

In this paper we show that the higher-order QNMs are also prominent in
binary BH mergers and they are interesting to search for in the
observations and simulations. Here the ``higher-order'' is with respect to 
the metric perturbations. Although the dimensionless amplitude of the metric
perturbations are negligible small when we observe them at detectors, they
are relatively large near the BH, up to 
\beqa
\psi^{(1)}/M\sim \left(E/M\right)^{1/2} \sim 10\%
\eeqa
for QNMs with energy $E/M \sim 1\%$ 
[see Eq.~(\ref{eq:E1}) with $\omega \sim M^{-1}$. 
$\psi^{(1)}$ denotes a wave function to be discussed later].
Hence the generated
second-order perturbations would have the amplitude $\sim 10\%$ of the 
first-order ones, which may be detectable for high SNR events.

Higher-order QNMs are essentially analogous to the anharmonic oscillations
discussed by Landau \& Lifshitz in ``Mechanics''~\cite{landau76}. In
general, an oscillation with small amplitude $x$ is described by an
equation,
$\ddot x+\omega^2 x=0$,
which gives the first-order solution $x=a \cos(\omega t + \phi)$. 
Here $a$ and $\phi$ are integration constants. 
Including the second-order term with respect to the amplitude $\sim x^2$,
the equation becomes
\beqa
\ddot x+\omega^2 x=-\alpha x^2 \,.
\eeqa
With the right hand side as a source term, we obtain a successive solution 
$x=a \cos(\omega t +\phi)+x^{(2)}$ where
\begin{eqnarray}
x^{(2)}=\frac{\alpha a^2}{6 \omega^2} \cos 2\omega t 
- \frac{\alpha a^2}{2\omega^2} \propto a^2 \,.
\end{eqnarray}
Here an oscillation arises with a frequency $2\omega$.
The important point is that the
second-order oscillation $x^{(2)}$ is driven by the first-order one
and thereby always exists.

Our result is that the second-order QNMs would also have
frequencies twice the first-order ones
$\omega^{(2)}=2 \omega^{(1)}$ 
and the amplitude up to $\sim 10\%$ 
of the first-order ones.
Since higher-order QNMs always exist, we may test
the nonlinearity of general relativity.
The purpose of this paper is 
to outline the calculations of the second-order QNMs for the
Schwarzschild BH and clarify the order-of-magnitude estimates. 
More details will be given in the forthcoming papers.
We use the units $c=G=1$, and
arbitrarily set $M=1$ which we can always recover, if necessary. 

Although this paper is the first to study second-order QNMs,
the second-order analysis is pioneered by Tomita~\cite{Tomita},
and the $\ell=2$, $m=0$ case in the Schwarzschild spacetime
is studied by Gleiser et al.~\cite{Gleiser:1995gx}.
It is also extended to cosmology~\cite{Acquaviva:2002ud} 
and the Kerr case~\cite{Campanelli:1998jv}. 

\section{First-order}
Let us consider the metric perturbations to second-order,
\beqa
\tilde g_{\mu \nu}=g_{\mu \nu}+h_{\mu \nu}^{(1)}+h_{\mu \nu}^{(2)} \,,
\eeqa
where $g_{\mu \nu}$ is the Schwarzschild metric and the
superscripts denote a perturbative order. Expanding the Einstein's 
vacuum equation we can obtain basic equations order 
by order~\cite{Gleiser:1995gx,Bruni:1996im}.

For the first-order, we use the Regge-Wheeler-Zerilli 
formalism~\cite{rw57,z70}.
Separating angular variables with tensor harmonics of indices $(\ell,m)$,
the equations decouple to the even (or polar) part with parity
$(-1)^\ell$ 
under a transformation $(\theta,\phi) \to (\pi-\theta,\pi+\phi)$
and 
the odd (or axial) part with parity $(-1)^{\ell+1}$.
Seven equations for the even parity part are reduced to a single Zerilli
equation,
and the other three equations for the odd parity part to a single Regge-Wheeler equation.
The Zerilli equation is given by
\beqa
&&\left[-\frac{\partial^2}{\partial t^2}
+\frac{\partial^2}{\partial r_*^2}
-V_{Z}(r)\right] \psi^{(1)}_{\ell m}(t,r)=0 \,;
\\
&&V_{Z}(r)=\left(1-\frac{2}{r}\right)
\frac{2\lambda^2 (\lambda+1) r^3+6 \lambda^2 r^2+
18 \lambda r +18}{r^3 (\lambda r+3)^2} \,,
\nonumber 
\eeqa
where $r_*=r+2 \ln \left(r/2-1\right)$ and
$\lambda=(\ell-1)(\ell+2)/2$. 
By Fourier transforming, 
$\psi^{(1)}_{\ell m}(t,r)=\int e^{-i\omega t} \psi^{(1)}_{\ell m
\omega}(r) d\omega$, the Zerilli equation gives a one-dimensional
scattering problem, which is familiar from quantum mechanics. The QNMs are
obtained by imposing the boundary conditions with purely ingoing waves
$\psi^{(1)}_{\ell m \omega} \sim e^{-i\omega r_*}$ at the horizon and purely
outgoing waves $\psi^{(1)}_{\ell m \omega} \sim e^{i\omega r_*}$ at
infinity. Such boundary conditions are
satisfied at discrete QNM frequencies $\omega^{(1)}_{\ell m n}$ 
that are complex with the
imaginary part representing the damping.
There is an infinite number of QNMs for each harmonics 
$(\ell,m)$~\cite{leaver85}.

All physical quantities for the first-order even parity part can be
reconstructed from $\psi^{(1)}_{\ell m}$. 
If we define 
\beqa
\psi^{(1)}_{\ell m}=\frac{r}{\lambda+1}
\left[K^{(1)}_{\ell m}+\frac{r-2M}{\lambda r+3M}
\left(H_{2\ell m}^{(1)}
-r \frac{\partial K_{\ell m}^{(1)}}{\partial r}\right)\right]
\eeqa
in the Regge-Wheeler (RW) gauge according to~\cite{Sago:2002fe,Martel:2005ir},
the GW power is given by
\beqa
\frac{dE^{(1)}}{dt}=\frac{1}{64\pi}\sum_{\ell m}
\frac{(\ell+2)!}{(\ell-2)!}
\left| \frac{\partial}{\partial t} \psi^{(1)}_{\ell m}(t,r) \right|^2 \,.
\label{eq:dEdt}
\eeqa
Note that in the RW gauge the gauge freedom is completely fixed
and the physical quantities can be expressed by
the gauge invariant functions in simple differential forms.

In this paper we focus on the most dominant modes in binary BH mergers,
i.e., the $\ell=2, m=2$ even parity mode for the first-order~\cite{Pretorius:2005gq}
and its driving second-order mode (the $\ell=4, m=4$ even parity mode as shown below).
Note that we have to specify $m$
since the degeneracy between $m$ breaks
at the second-order.

\section{Second-order}
For the second-order, we also separate angular
variables in terms of tensor harmonics. 
Instead of $\psi^{(2)}_{\ell m}$, 
we introduce a function
\begin{eqnarray}
\chi^{(2)}_{\ell m}=\frac{r-2M}{\lambda r+3M}
\left[\frac{r^2}{r-2M} 
\frac{\partial K^{(2)}_{\ell m}}{\partial t} - H_{1\ell m}^{(2)}\right]
\label{eq:defchi}
\end{eqnarray}
in the RW gauge for convenience, which is essentially the time-derivative of 
$\psi^{(2)}_{\ell m}$~\cite{Gleiser:1995gx}.
The first-order counterpart exactly
satisfies $\chi^{(1)}_{\ell m}=\partial
\psi^{(1)}_{\ell m}/\partial t$, and so the dimensions are
$\psi^{(i)}_{\ell m} \sim O(M)$ and 
$\chi^{(i)}_{\ell m} \sim O(M^0)$ ($i=1,\,2$).
The equations for the even parity part are reduced to the Zerilli equation with a
source term,
\begin{eqnarray}
\left[-\frac{\partial^2}{\partial t^2}
+\frac{\partial^2}{\partial r_*^2}
-V_{Z}(r)\right] \chi^{(2)}_{\ell m}(t,r)=S_{\ell m}(t,r) \,,
\label{eq:Zeqchi}
\end{eqnarray}
where the source term $S_{\ell m}$ is quadratic in $\psi^{(1)}_{\ell m}$.

The most dominant second-order mode is the $\ell=4$, $m=4$ even parity one.
This is because the dominant first-order mode is the 
$\ell=2$, $m=2$ even parity mode and hence
$S_{\ell m}$ is dominated by the product
$(\ell=2,m=2) \times (\ell=2,m=2)$ in Eq.~(\ref{eq:Zeqchi}).
This gives the $m=4$ mode that is $\ell=4$ and even parity.
Note that what we are considering is the particular solution.
Homogeneous solutions are just proportional to
the first-order ones and are therefore trivial.

\section{Regularization}
Although it is straightforward to calculate the
$\ell=4$, $m=4$ source term $S_{44}$ in terms of
$\psi^{(1)}_{22}$, the raw source term does not behave
well at infinity and is not suitable for calculations. We can
find $S_{44} \sim O(r^0)$ at infinity by using
the expansion
\begin{eqnarray}
\psi^{(1)}_{22}= 
\frac{1}{3}F_I''+\frac{1}{r}F_I'
+\frac{1}{r^2}\left(F_I-F_I'\right)+O(r^{-3}) \,,
\end{eqnarray}
where $F_I=F_I(T_{-})$ is some function of
$T_{-}=t-r_*$ and $F_I'$ denotes 
$dF_I(T_{-})/dT_{-}$. 
We need a regularized source $S_{44}^{reg} \sim O(r^{-2})$ at least 
[i.e., the same as the
potential $V_Z \sim O(r^{-2})$]. At the horizon
the source behaves well, $S_{44} \sim O[(r-2)]$, with
\begin{eqnarray}
\psi^{(1)}_{22}=F_H'+\frac{1}{4}F_H
+\frac{27}{56}(r-2)F_H
 +O[(r-2)^2] \,,
\end{eqnarray}
where $F_H=F_H(T_{+})$ is some function of $T_{+}=t+r_*$.

We can regularize the source term by using the regularized function,
\begin{eqnarray}
\chi^{(2) reg}_{44}
= \chi^{(2)}_{44}
-\frac{\sqrt{70}}{126\sqrt{\pi}}
\frac{(r-2)^2}{r}
\frac{\partial \psi^{(1)}_{22}}{\partial r}
\frac{\partial^2 \psi^{(1)}_{22}}{\partial r \partial t} \,,
\label{eq:chireg}
\end{eqnarray}
which satisfies the Zerilli equation (\ref{eq:Zeqchi})
with a well-behaved source term,
$S_{44}^{reg}\sim O(r^{-2})$ at infinity and 
$S_{44}^{reg}\sim O[(r-2)]$ at the horizon.
Thus we can remove an unphysical gauge-dependent divergence.
Note that such a regularization is not unique, 
and for example we can replace $\partial/\partial r$ with 
$\partial/\partial t$ in Eq.~(\ref{eq:chireg}).
The regularization is equivalent to
adding quadratic terms of the first-order gauge invariant
function to the second-order gauge invariant function,
so that it preserves the gauge 
invariance~\cite{Garat:1999vr}.

The explicit form of the regularized source term is
\begin{widetext}
\begin{eqnarray}
S^{reg}_{44}(t,r) &=&
\frac{r-2}{42} \frac{\sqrt{70}}{\sqrt {\pi }}
\Biggl\{  
{\frac { 228{r}^{7}+8{r}^{6}-370{r}^{5}+142{r}^{4}-384{r}^{3}
- 514{r}^{2}-273r-48 }
{{r}^{5} \left( 3r+1 \right) ^{2} \left( 2r+3 \right) ^{2}}}
\dot \psi'
\psi'
\cr
&-& {\frac {72{r}^{8}+3936{r}^{7}+2316{r}^{6}
- 2030{r}^{5}
-7744{r}^{4}-9512{r}^{3}-3540{r}^{2}
-1119r-144}
{{r}^{6} \left( 3r+1 \right) ^{2} 
\left( 2r+3 \right) ^{3}}}
\psi'
\dot \psi
\cr
&+& {\frac { \left( 7r+4 \right) r  }{3 \left( r-2 \right) ^{2}}}
\dddot \psi
\ddot \psi
+{\frac {24{r}^{7}+344{r}^{6}-872{r}^{5}-771{r}^{4}
+120{r}^{3}+77{r}^{2}-237r-48
}
{{r}^{3} \left( 3r+1
 \right) ^{2} \left( 2r+3 \right) ^{2} \left( r-2 \right) ^{2}}}
\ddot \psi
\dot \psi
\cr
&-& {\frac { 66{r}^{4}-106{r}^{3}-220{r}^{2}-156r-45 }
{r \left( 3r+1
 \right) ^{2} \left( 2r+3 \right)  \left( r-2 \right) }}
\ddot \psi'
\dot \psi
+{\frac { 
198{r}^{5}-318{r}^{4}-664{r}^{3}-458{r}^{2}-127r-24 }
{3{r}^{2} \left( 3r+1 \right) ^{2}
 \left( 2r+3 \right)  \left( r-2 \right) }} 
{\dddot{\psi}}
\psi'
\cr
&-& \frac {3 \left(2160{r}^{9}+11760{r}^{8}
+30560{r}^{7}+41124{r}^{6}
+ 31596{r}^{5}+11630{r}^{4}
-1296{r}^{3}-4182{r}^{2}-1341r-144 \right) }
{{r}^{7} \left( 2r+3 \right) ^{4} \left( 3r+1 \right) ^{2}}
\psi 
\dot \psi
\cr
&-& {\frac { 7r+4 }{3r}}
\ddot \psi'
\dot \psi'
- {\frac {
 2 \left( r-2 \right) }{3 \left( 3r+1 \right) ^{2}{r}^{2}}}
\ddot \psi
\dot \psi'
+ {\frac { 252{r}^{6}+636{r}^{5}+674{r}^{4}
+730{r}^{3}+524{r}^{2}+171r+24 }
{{r}^{3} \left( 3r+1 \right) ^{2}
 \left( 2r+3 \right) ^{2} \left( r-2 \right) }} 
\psi 
\dddot \psi
\cr
&-& {\frac { 216{r}^{8}+4296{r}^{7}
+1992{r}^{6}
-3488{r}^{5}
- 8716{r}^{4}-9512{r}^{3} 
-3540{r}^{2}-1119r
-144 }
{{r}^{6} \left( 3r+1 \right) ^{2} \left( 2r+3 \right) ^{3}}} 
\psi 
\dot \psi'
\Biggr\} \,,
\label{eq:Sreg}
\end{eqnarray}
\end{widetext}
where $\dot \psi=\partial \psi^{(1)}_{22}/\partial t$ and 
$\psi'=\partial \psi^{(1)}_{22}/\partial r$. 
We can now solve the regularized Zerilli equation~(\ref{eq:Zeqchi}). 
With Fourier expansions, this provides a two-point
boundary value problem with purely ingoing boundary condition 
at the horizon and purely outgoing at infinity.
The numerical calculation will be presented 
in a forthcoming paper.

\section{Detectability}
Without solving Eq.~(\ref{eq:Zeqchi}) numerically,
we can find the essential properties of the solutions, 
i.e., the QNM frequency $\omega^{(2)}_{44 n}$
and the order-of-magnitude QNM amplitude. 
Since the source term $S^{(2) reg}_{44}$ 
is quadratic in the first-order function 
$\psi^{(1)}_{22} \propto e^{-i\omega^{(1)}_{22 n} t}$,
we have $\chi_{44}^{(2) reg} \propto e^{-2 i\omega^{(1)}_{22 n} t}$
from Eq.~(\ref{eq:Zeqchi}).
Therefore the second-order QNM frequencies are
twice the first-order ones,
\beqa
\omega^{(2)}_{44 n}=2 \omega^{(1)}_{22 n} \,,
\eeqa
which differ from any first-order QNM
frequencies $\omega^{(1)}_{\ell m n}$.
By matching the dimensions in both sides of Eq.~(\ref{eq:Zeqchi}),
we also estimate the order-of-magnitude amplitude as
\beqa
\left|\chi^{(2) reg}_{44}\right| 
\sim \left|\omega^{(1)}_{22 n} \psi^{(1)}_{22} \right|^2 \,.
\label{eq:ampli}
\eeqa
Then we can derive the second-order QNM energy $E^{(2)}$
from the first-order one $E^{(1)}$.
For the QNM waveform, $\psi^{(1)}_{22}=\psi^{(1)}_{22}(0) 
e^{-i\omega^{(1)}_{22 n} t}$,
the first-order energy $E^{(1)}$ in Eq.~(\ref{eq:dEdt}) is given by
\beqa
\frac{E^{(1)}}{M} = \frac{3}{8\pi} 
\frac{\left|\omega^{(1)}_{22 n}\right|^2 
\left|\psi^{(1)}_{22}(0)\right|^2}
{M \left|\Im \omega^{(1)}_{22 n}\right|} \,,
\label{eq:E1}
\eeqa
where $E^{(1)}/M \sim 1\%$ 
for equal-mass mergers \cite{Pretorius:2005gq}.
By noting $\chi^{(2)}_{44} \sim \partial \psi^{(2)}_{44}/\partial t 
\sim 2 \omega^{(1)}_{22 n} \psi^{(2)}_{44}$,
we have a similar expression for the second-order energy $E^{(2)}$ as
\beqa
\frac{E^{(2)}}{M} 
\sim \frac{45}{8\pi}
\frac{\left|\chi^{(2) reg}_{44}(0)\right|^2}
{M \left|2 \Im\omega^{(1)}_{22 n}\right|}
\sim M |\Im \omega^{(1)}_{22 n}|
\left(\frac{E^{(1)}}{M}\right)^2 \,,
\label{eq:E2}
\eeqa
where the second equality uses Eqs.~(\ref{eq:ampli}) and (\ref{eq:E1}).
In a next paper we will calculate
the coefficient as $E^{(2)}/M \sim 15 M |\Im \omega^{(1)}_{22 n}|
(E^{(1)}/M)^2$.

Once we know the QNM frequency $\omega$ and energy $E$,
we can obtain the GW energy 
spectrum~\cite{Flanagan:1997sx,Berti:2005ys}
\beqa
\frac{dE}{df} 
\simeq \frac{16\pi^2 E f^2 |\Im \omega|^3}
{|\omega|^2 \left[(2\pi f-\Re \omega)^2+(\Im \omega)^2\right]^2} \,,
\label{eq:dEdf}
\eeqa
and then the characteristic amplitude $h_{char}(f)$ 
with Eq.~(5.1) of Flanagan and Hughes \cite{Flanagan:1997sx},
\beqa
h_{char}(f)^2 = \frac{2(1+z)^2}{\pi^2 D(z)^2} \frac{dE}{df}[(1+z)f] \,,
\label{eq:hchar}
\eeqa
where $D(z)$ is the luminosity distance.

\section{Discussions}
\begin{figure}
  \begin{center}
\epsfxsize=8.cm
\begin{minipage}{\epsfxsize} 
\epsffile{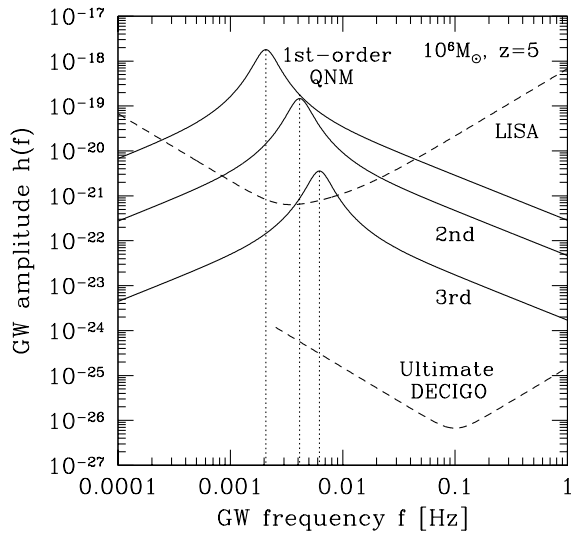} 
\end{minipage}
  \caption{
The characteristic GW amplitudes $h_{char}(f)$
of the Schwarzschild black hole (BH) 
quasi-normal modes (QNMs)
for two equal-mass binary BH mergers 
of total mass $10^6 M_{\odot}$ at redshift $z=5$.
The first-order QNM has $\ell=2, m=2$ and energy $E^{(1)}/M=1\%$.
The second and third-order QNMs also appear
at twice and three times the first-order frequency,
respectively, with detectable amplitudes.
The dashed line is the rms noise amplitude 
$h_n(f)=\sqrt{5 f S_h(f)}$ for 
the space-based detector LISA and ultimate DECIGO.
The signal-to-noise ratio squared for a randomly
oriented source is given by 
$(SNR)^2=\int d(\ln f)[h_{char}(f)/h_n(f)]^2$.
We use a cosmology $(\Omega_m,\Omega_\Lambda,h)=(0.27,0.73,0.71)$.
We estimate the third-order QNM energy 
by extrapolating the second-order equation (\ref{eq:E2})
as $E^{(3)}/M \sim 15 (M \Im \omega)^2 (E^{(1)}/M)^3$.
}
\label{fig:snr}
 \end{center}
\end{figure}

Our main results are summarized in Fig.~\ref{fig:snr} showing the
characteristic GW amplitudes $h_{char}(f)$ of the Schwarzschild BH
QNMs for two equal-mass binary BH mergers
of total mass $10^6 M_{\odot}$ at redshift $z=5$, together with the rms
noise amplitude $h_n(f)=\sqrt{5 f S_h(f)}$ for the space-based detector 
LISA~\cite{Flanagan:1997sx,Berti:2005ys}
and ultimate DECIGO~\cite{Seto:2001qf}. The first-order QNM has $\ell=2,
m=2$ and energy $E^{(1)}/M=1\%$. We estimate the second-order QNM with
Eqs.~(\ref{eq:E2})--(\ref{eq:hchar}) and the third-order QNM energy by
extrapolating the second-order equation (\ref{eq:E2}) as $E^{(3)}/M \sim
15 (M \Im \omega)^2 (E^{(1)}/M)^3$. We can see that the second and
third-order QNMs appear at twice and three times the first-order
frequency, respectively, with detectable amplitudes.

We can in principle identify the higher-order QNMs
since their frequencies differ from any first-order ones.
However the actual identification
depends on the SNR of the observations~\cite{Berti:2005ys}
or the accuracy of the simulations.
In simulations with current accuracy
we may have already mistaken the second-order QNMs for
the first-order ones.
For example $2 M \omega^{(1)}_{2m0}=0.7473-0.1779i$ is close to
$M \omega^{(1)}_{4m0}=0.80918-0.09416 i$~\cite{leaver85}.

In order to proof that the second-order QNMs actually exist,
we have to find the second-order QNMs directly
in the numerical simulations.
Such simulations are challenging because the mesh size
should be less than $\sim 1\% \times M$
to resolve $\sim 1\%$ metric perturbations.
Even 1+1D spherical models for the fully self-gravitating case 
have not found the second-order QNMs~\cite{Zlochower:2003yh}.
Simulations of acoustic black holes may be an alternative
for this purpose~\cite{Okuzumi:2007hf}.
We also need a mathematically rigorous definition of second-order
QNMs like the first-order ones that use the Laplace transformation
rather than the Fourier transformation~\cite{leaver86,Kokkotas:1999bd}.

Future problems include the Kerr BH case, the odd parity mode case,
and a more solid third-order formulation.
Since the master equation for Kerr BHs also has a source term 
quadratic in the first-order function~\cite{Campanelli:1998jv}, 
we may expect similar results. When BHs have no spin before mergers,
the final spin is $a \sim 0.7$~\cite{Pretorius:2005gq}
and hence the Kerr effects may not be so large
(as inferred from the fact that the QNM frequencies shift 
by only a small factor).
The odd parity mode appears when BHs have spin before mergers.
At the third-order the QNM frequencies will also shift up to
$(\psi^{(1)}/M)^2\sim 1\%$
as suggested by the anharmonic oscillations in~\cite{landau76},
probably blueward because the GW carries away the BH mass.

The ratio between the first and higher-order QNM amplitude
include new information about the total GW energy $E$.
Since the observed GW amplitude is $h \sim (E/M)^{1/2}(M/r)$,
this could provide a distance indicator.

\acknowledgments
We thank C.~O.~Lousto and Y.~Zlochower for a careful reading of the manuscript.
This work is supported in part by
the Grant-in-Aid (18740147) from the 
Ministry of Education, Culture, Sports, Science and Technology
(MEXT) of Japan (KI). 
This is also supported by JSPS Postdoctoral Fellowships 
for Research Abroad and in part by the NSF 
for financial support from grant PHY-0722315 (HN).



\end{document}